\documentclass[letterpaper, 10 pt, conference]{ieeeconf}

\IEEEoverridecommandlockouts

\usepackage[dvipsnames]{xcolor}   

\usepackage{cite}

\usepackage{graphicx}   
\usepackage[font=footnotesize, labelsep=period]{caption}
\usepackage[font=footnotesize, labelsep=period]{subcaption} 

\let\labelindent\relax
\usepackage{enumitem}

\usepackage{amsmath,amssymb,amsfonts}
\usepackage{siunitx}

\usepackage{algorithm}
\usepackage[noend]{algpseudocode}

\usepackage{booktabs}   
\usepackage{multirow}   
\usepackage{array}

\usepackage[hidelinks]{hyperref}
\usepackage{url}

\usepackage{orcidlink}

\usepackage{balance}  
\usepackage[
    colorinlistoftodos,  
    textsize=small,      
    backgroundcolor=yellow!20 
]{todonotes}
\usepackage{xparse} 

\usepackage{amsthm}  

\theoremstyle{definition} 
\newtheorem{definition}{Definition}

\theoremstyle{remark} 
\newtheorem{remark}{Remark}

\newcommand{\q}{\mathbf{q}}         
\newcommand{\qe}{\boldsymbol{\eta}} 

\newcommand{\qA}{\mathbf{q}^{\mathrm{A}}}

\newcommand{\qD}{\mathbf{q}^{\mathrm{D}}}

\newcommand{\qDbar}{\bar{\mathbf{q}}^{\mathrm{D}}}

\newcommand{\x}{\mathbf{x}}
\newcommand{\xD}{\mathbf{x}^{\mathrm{D}}}

\newcommand{\xA}{\mathbf{x}^{\mathrm{A}}}
\newcommand{\xAk}[1]{\mathbf{x}_{#1}^{\mathrm{A}}}
\newcommand{\xe}{\boldsymbol{\chi}}
\newcommand{\xeD}{\boldsymbol{\chi}^{\mathrm{D}}}

\newcommand{\xeA}{\boldsymbol{\chi}^{\mathrm{A}}}

\newcommand{\dD}{\boldsymbol{\delta}^{\mathrm{D}}}

\newcommand{\dA}{\boldsymbol{\delta}^{\mathrm{A}}}

\newcommand{\z}{\mathbf{z}}
\newcommand{\zD}{\mathbf{z}^{\mathrm{D}}}
\newcommand{\zDk}[1]{\mathbf{z}_{#1}^{\mathrm{D}}}
\newcommand{\zA}{\mathbf{z}^{\mathrm{A}}}

\let\u\relax
\newcommand{\u}{\mathbf{u}}

\newcommand{\uD}{\mathbf{u}^{\mathrm{D}}}

\newcommand{\uA}{\mathbf{u}^{\mathrm{A}}}

\newcommand{\gammaD}{\boldsymbol{\gamma}^{\mathrm{D}}}
\newcommand{\gammaA}{\boldsymbol{\gamma}^{\mathrm{A}}}

\title{\LARGE \bf
Regulation-Aware Game-Theoretic Motion Planning for \\ Autonomous Racing
}

\author{%
    Francesco Prignoli$^{1,2}$, Francesco Borrelli$^{3}$, Paolo Falcone$^{1,4}$, and Mark Pustilnik$^{3}$%
    \thanks{$^{1}$Francesco Prignoli and Paolo Falcone are with the "Enzo Ferrari" Department of Engineering (DIEF),
        University of Modena and Reggio Emilia, 41125 Modena, Italy
        {\tt\small \{francesco.prignoli, paolo.falcone\}@unimore.it}}%
    \thanks{$^{2}$Francesco Prignoli is also with the Department of Electrical, Electronic and Information Engineering “Guglielmo Marconi” (DEI),
        University of Bologna, 40136 Bologna, Italy
        {\tt\small francesco.prignoli@unibo.it}}%
    \thanks{$^{3}$Francesco Borrelli and Mark Pustilnik are with the Department of Mechanical Engineering, University of California, Berkeley, CA 94720, USA
        {\tt\small \{fborrelli,pkmark\}@berkeley.edu}}%
    \thanks{$^{4}$Paolo Falcone is also with the Electrical Engineering Department, Chalmers University of Technology, Gothenburg, Sweden
        {\tt\small falcone@chalmers.se}}%
}

\begin{document}

\maketitle
\thispagestyle{empty}
\pagestyle{empty}

\begin{abstract}
This paper presents a regulation-aware motion planning framework for autonomous racing scenarios.  
Each agent solves a Regulation-Compliant Model Predictive Control problem, where racing rules---such as right-of-way and collision avoidance responsibilities---are encoded using Mixed Logical Dynamical constraints.
We formalize the interaction between vehicles as a Generalized Nash Equilibrium Problem (GNEP) and approximate its solution using an Iterative Best Response scheme.  
Building on this, we introduce the Regulation-Aware Game-Theoretic Planner (RA-GTP), in which the attacker reasons over the defender’s regulation-constrained behavior.  
This game-theoretic layer enables the generation of overtaking strategies that are both safe and non-conservative. 
Simulation results demonstrate that the RA-GTP outperforms baseline methods that assume non-interacting or rule-agnostic opponent models, leading to more effective maneuvers while consistently maintaining compliance with racing regulations.
\end{abstract}

\section{Introduction}
Autonomous racing has emerged as an important accelerator for advanced motion planning and control algorithms, pushing the boundaries of autonomous navigation in highly-dynamic and safety-critical scenarios.
In particular, multi-agents competitions drive research towards trajectory planning methods that aim to achieve optimal goal-reaching behaviors while safely navigating complex interactions with other agents.
Notably, traditional predict-then-plan approaches often lead to overly conservative maneuvers, resulting in passive behaviors that fail to account for how other agents might react during the planning phase. To overcome these limitations, game-theoretic approaches have been proposed, aiming to devise strategies that explicitly consider and potentially leverage the influence of the planned trajectory on other agents.

In full-scale autonomous racing competitions such as the Indy Autonomous Challenge\footnote{\url{https://www.indyautonomouschallenge.com/}} and the Abu Dhabi Autonomous Racing League\footnote{\url{https://a2rl.io/}}, trajectory-sampling-based planning methods have been widely adopted due to their flexibility and the ease of integrating post-evaluation checks---primarily for collision avoidance, but not limited to it~\cite{raji_erautopilot_2024,chung_autonomous_2024}. However, this explicit assessment typically comes at the cost of coarse sampling resolution, which can result in suboptimal trajectories. Moreover, while these approaches can handle static rule sets effectively, they often struggle to incorporate interaction-based regulations---such as right-of-way rules---where the planning outcome depends on the dynamic behavior and intent of other agents on the track.
In addition, regulation-compliant approaches are typically limited to regulation assessment\cite{ogretmen_hybrid_2023}, and are not exploited to enhance the effectiveness of overtaking strategies .
In reinforcement learning (RL)-based approaches \cite{song_autonomous_2021}, such strategic exploitation of regulations may be implicitly embedded within the learned policy. However, these methods generally lack formal guarantees of compliance, limiting their applicability in safety-critical racing scenarios.

By modeling vehicle interactions in head-to-head racing as a non-cooperative game, different competitive behaviors can be captured by analyzing either Nash or Stackelberg equilibria~\cite{liniger_noncooperative_2020}. When agents are subject to coupling constraints such as collision avoidance, the solution concept extends to Generalized Nash Equilibria (GNE)~\cite{facchinei_generalized_2010}. In such settings, Iterative Best Response (IBR) algorithms has been adopted to approximate GNE solutions~\cite{williams_best_2018}. IBR methods guarantee that their fixed points correspond to GNE under certain assumptions~\cite{facchinei_decomposition_2011}. However, in dynamic games, multiple Nash equilibria may exist, and the specific equilibrium reached by IBR often depends on the initial conditions used in the iteration.  While each agent optimizes its individual objective given the strategies of the others, there is generally no notion of global optimality in the multi-agent sense. Instead, optimality is assessed from the perspective of each individual agent. To address this ambiguity, the works in~\cite{wang_game-theoretic_2021,spica_real-time_2020} introduces a sensitivity term that influences how the joint strategy converges, effectively allowing a designer to steer the final equilibrium toward more favorable outcomes for a given agent. This capability is especially relevant when the agents have symmetric objectives or constraints, as such symmetry can otherwise lead to equilibria in which no agent’s preference dominates, and small changes in initial conditions can yield drastically different equilibrium outcomes.
When all the coupling constraints are shared among agents, the problem reduces to finding a non-normalized solution of a GNE~\cite{pustilnik_non-normalized_2025}.
However, in real-world scenarios, it is precisely asymmetry that grants one agent an advantage over others. Such asymmetry can arise from differences in vehicle performance or varying risk tolerances among competitors. Crucially, in organized racing, regulations often amplify these imbalances intentionally---not only to prohibit unsafe maneuvers, but also to increase entertainment value by creating strategic inequities among participants. This highlights the importance of considering regulatory compliance when designing planning algorithms, ensuring natural behaviors and competitive realism.

This work addresses the challenge of enforcing racing rules in autonomous motion planning to reduce overly conservative behavior and promote competitive yet rule-compliant interactions with opponents. Motivated by real-world autonomous racing competitions, we focus on the enforcement of \emph{right-of-way} rules, which restrict the lateral motion of the leading vehicle during overtaking maneuvers. To formally encode these rules within an optimization-based planner, we adopt the Mixed Logical Dynamical (MLD) framework, which enables the integration of logical decisions and continuous vehicle dynamics into a unified predictive control scheme~\cite{bemporad_control_1999}.
The key contributions of this work are threefold:  
(1) a formal mathematical description of the overtaking regulations adopted in autonomous racing;  
(2) the development of a Regulation-Compliant MPC scheme that enforces overtaking rules using the MLD modeling framework; and  
(3) the design of a Regulation-Aware Game-Theoretic Planner that reasons over the opponent behavior to plan effective overtaking strategies.  
The proposed planner outperforms standard approaches that assume fixed opponent predictions, while eliminating the need for auxiliary sensitivity-based manipulations.

The rest of the paper is structured as follows: in Sec.~\ref{sec:problem-settings} we present the problem setting. In Sec.~\ref{sec:right-of-way-encoding}, we introduce a formal definition of the overtaking regulation. In Sec.~\ref{sec:regulation-compliant-MPC}, we incorporate these regulatory constraints into a Regulation-Compliant MPC scheme. In Sec.~\ref{sec:regulation-aware-GTMP}, we illustrate how an agent can leverage the right-of-way regulation to plan effective overtaking maneuvers by proposing a game-theoretic motion planner. Finally, in Sec.~\ref{sec:simulation} we show the results of simulation studies comparing the traditional fixed prediction approach against ours.

\section{Problem Setting} \label{sec:problem-settings}

We consider a bounded, planar track and a reference path --- the racing line --- defined as an arc-length parameterized curve 
\( \boldsymbol{\xi}: [0, S_{\max}] \rightarrow \mathbb{R}^2\), assumed to be at least twice continuously differentiable (i.e., \(\boldsymbol{\xi}(s) \in C^2\)). Let \(s \in [0, S_{\max}]\) denote the arc length parameter along this curve. The corresponding curvature profile is given by a scalar function \( \kappa: [0, S_{\max}] \rightarrow \mathbb{R}\), which is assumed to be continuous.
In the Frenet frame defined w.r.t. \(\boldsymbol{\xi}(s)\), the left and right track boundaries are described by two continuous functions 
\(\beta_{\ell}(s),\; \beta_{r}(s): [0, S_{\max}] \rightarrow \mathbb{R}\), respectively, defined as the signed lateral distances from \(\boldsymbol{\xi}(s)\).
These offsets are not constant and may vary along the track, as the racing line does not, in general, coincide with the geometric centerline.
Let \(\q \in \mathcal{Q} \subseteq \mathbb{R}^2\) denote the position of the vehicle’s center of gravity (CoG) in the Frenet frame, with \(\q \triangleq \begin{bmatrix} s & n\end{bmatrix}^{\top}\), where \(s \in [0,S_{\max}]\) is the arc length along the racing line, and \(n \in \mathbb{R}\) is the signed lateral displacement from \(\boldsymbol{\xi}(s)\). The set 
\begin{equation}
    \mathcal{Q} \triangleq \left\{\q \in \mathbb{R}^2 \mid n_{r}(s) \leq n \leq n_{\ell}(s)\right\}
\end{equation} defines the admissible positions such that the vehicle is within the track boundaries. Here, the functions \(n_{\ell}(s),\; n_{r}(s) : [0, S_{\max}] \rightarrow \mathbb{R}\) are derived by tightening the geometric track boundaries \(\beta_{\ell}(s)\) and \(\beta_r(s)\) by a fixed margin that accounts for the vehicle’s physical dimensions, the curvature of the racing line, and the expected vehicle's maximum heading error.

\section{Racing Regulation Modeling} \label{sec:right-of-way-encoding}
In this work, we consider a head-to-head autonomous racing scenario in which we refer to the vehicle attempting to overtake as the \emph{attacker} and the vehicle being overtaken as the \emph{defender}. Inspired by real-world autonomous racing competitions, we focus on the regulation that constrains the vehicles’ motion during an overtaking maneuver.

\begin{figure}[t]
    \centering
    \includegraphics[width=\linewidth]{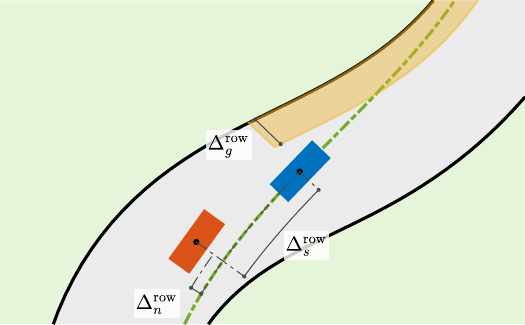}
    \caption{Illustration of the attacker (orange) acquiring the right of way per rules~\ref{def:row-verbal}. The defender (blue) must yield the yellow shaded region.}
    \label{fig:row-picture}
\end{figure}

\begin{definition}[Overtaking Regulation]\label{def:row-verbal}
    The \emph{Overtaking Regulation} defines the rules for the attacker (A) and the defender (D) during an overtaking maneuver (see Fig.~\ref{fig:row-picture}):
    \begin{enumerate}[label=(R\arabic*)]
        \item \textbf{Right-of-Way Acquisition:} The attacker gains the right-of-way once its relative longitudinal distance to the defender is smaller than a prescribed distance \(\Delta_s^{\mathrm{row}}\) and their lateral separation is above a prescribed margin \(\Delta_n^{\mathrm{row}}\), which also determines the overtaking side.
        \item \textbf{Yielding Obligation:} Once the attacker gains the right-of-way, the defender must provide sufficient lateral space \(\Delta_g^{\mathrm{row}}\) on the overtaking side, provided that such space was available at the moment of the right-of-way acquisition.
        \item \textbf{Collision Avoidance Responsibility:} Throughout the overtaking maneuver, the attacker is responsible for avoiding collisions.
    \end{enumerate}
\end{definition}
\begin{remark}
The proposed set of rules offers a simplified yet practical abstraction of overtaking regulation.  
Different parameter choices can reflect more restrictive or permissive interpretations of right-of-way.  
A detailed modeling of competition-specific regulations is beyond the scope of this work.
\end{remark}
Having formally defined the race regulations, we now translate them into a mathematical formulation in the form of conditional constraints.
Define the joint vehicle position in the Frenet frame as \(\qe \triangleq \begin{bmatrix} \qD & \qA\end{bmatrix}^{\top} \in \mathcal{Q}^2 \), where \(\qD\) and \(\qA\) are the defender's and attacker's positions, respectively.
To characterize the relative longitudinal and lateral positioning between the two vehicles, we introduce the following four scalar functions \(h_* : \mathcal{Q}^2 \rightarrow \mathbb{R}\):
\begin{subequations}\label{eq:h-functions}
        \begin{align}
            h_{s^\pm}(\qe) & \triangleq \pm s_{\mathrm{D}} \mp s_{\mathrm{A}} - \Delta_s^{\mathrm{row}}, \label{eq:indicator_s+} \\
            h_{n^\pm}(\qe) & \triangleq \pm n_{\mathrm{D}} \mp n_{\mathrm{A}} + \Delta_n^{\mathrm{row}}, \label{eq:indicator_n+}
        \end{align}
\end{subequations}
along with the corresponding sets \(\mathcal{S}_* \subseteq \mathcal{Q}^2\):
\begin{equation}
    \mathcal{S}_* \triangleq \left\{ \qe \in \mathcal{Q}^2 \mid h_*(\qe) \leq 0 \right\} \quad \forall\, *\in \mathcal{I}_{sn} ,
    \label{eq:h sets}
\end{equation}
where \(\mathcal{I}_{sn} \triangleq \{s^+,s^-,n^+,n^-\}\).
Additionally, we define the set \(\mathcal{S}_{s} \triangleq \mathcal{S}_{s^+} \cap \mathcal{S}_{s^-}\). 

Define the joint \emph{crossing} position \( \bar{\qe} \triangleq \qe(\bar{t})\), with
\begin{equation} \label{eq:t-bar}
    \bar{t} \triangleq \sup \left\{ t \in \mathbb{R}^+ \mid h_{s^+}(\qe(t)) = 0 \right\},
\end{equation} 
that is the latest time the attacker closed the longitudinal gap to \(\Delta_s^{\mathrm{row}}\). 
Let \( \Delta_{\ell},\; \Delta_{r}: \mathcal{Q} \rightarrow \mathbb{R}\) be the signed lateral distance to the left and right tightened track margins, respectively:
\begin{equation}
    \begin{aligned}
        \Delta_{\ell}(\q) = n_{\ell}(s) - n, \\
        \Delta_{r}(\q) = n - n_{r}(s).
    \end{aligned}
\end{equation}
Additionally, let \(\Delta_{g\ell},\; \Delta_{gr}: \mathcal{Q} \rightarrow \mathbb{R}\) be the effective granted lateral space that the defender must leave on the left and right side, respectively, defined as:
\begin{equation}\label{eq:effective-granted-space}
    \begin{aligned}
         \Delta_{g\ell}(\q) & \triangleq \min\bigl(\Delta_g^{\mathrm{row}}, \Delta_{\ell}(\q) \bigr), \\
        \Delta_{gr}(\q) & \triangleq \min\bigl(\Delta_g^{\mathrm{row}}, \Delta_{r}(\q) \bigr).
    \end{aligned}
\end{equation}
Finally, we define the sets \( \mathcal{Q}_\ell^{\mathrm{row}},\, \mathcal{Q}_r^{\mathrm{row}} \subseteq \mathcal{Q} \) as the admissible defender positions corresponding to a given crossing position \( \bar{\q}_D \), for a left and right overtake, respectively:
\begin{equation}
    \begin{aligned}
        \mathcal{Q}_{\ell}^{\mathrm{row}}(\qDbar) \triangleq \{\qD \in \mathcal{Q} \mid h_{\ell}^{\mathrm{row}}(\qD, \qDbar) \leq 0 \}, \\
        \mathcal{Q}_{r}^{\mathrm{row}}(\qDbar) \triangleq \{\qD \in \mathcal{Q} \mid h_{r}^{\mathrm{row}}(\qD, \qDbar) \leq 0 \},
    \end{aligned}
    \label{eq:row-sets}
\end{equation}
where
\begin{equation}
    \begin{aligned}
        h_{\ell}^{\mathrm{row}}(\qD, \qDbar) & \triangleq -\Delta_{\ell}(\qD) + \Delta_{g\ell}(\qDbar),\\
        h_{r}^{\mathrm{row}}(\qD, \qDbar) & \triangleq -\Delta_{r}(\qD) + \Delta_{gr}(\qDbar).
    \end{aligned}
\end{equation}
In words, the sets in~\eqref{eq:row-sets} define the admissible defender positions obtained by reducing the lateral margins \(n_{\ell}\) and \(n_r\) by the effective granted space \(\Delta_{g\ell}\) and \(\Delta_{gr}\), each calculated as the smallest between the prescribed offset \(\Delta_g^{\mathrm{row}}\) and the available lateral clearance \(\Delta_{\ell}(\qDbar)\) and \(\Delta_{r}(\qDbar) \) at time~$\bar{t}$, respectively.

\begin{definition}[Right-of-Way Compliance]
The defender’s position \( \qD \in \mathcal{Q} \) is said to be \emph{right-of-way compliant} with respect to the joint current position \( \qe \) and the joint crossing position \( \bar{\qe} \) if the following conditions hold:
    \begin{equation}
    \left\{
    \begin{aligned}
    \text{if } \qe \in \mathcal{S}_s \text{ and } \bar{\qe} \in \mathcal{S}_{n^+}, 
    &\quad \text{ then } \qD \in \mathcal{Q}_{\ell}^{\mathrm{row}}(\qDbar), \\
    \text{if } \qe \in \mathcal{S}_s \text{ and } \bar{\qe} \in \mathcal{S}_{n^-}, 
    &\quad \text{ then } \qD \in \mathcal{Q}_{r}^{\mathrm{row}}(\qDbar).
    \end{aligned}
    \right.
    \label{eq:row-constr-conditional}
    \end{equation}
    \label{def:row constr-conditional}
\end{definition}
Formulation~\eqref{eq:row-constr-conditional} highlights both the conditional nature of the right‑of‑way constraints imposed on the defender and their dependence on the joint crossing position \(\bar{\qe}\). The remainder of this section discusses how to encode these constraints as inequalities, so they can be embedded in an optimization problem.

Let us introduce the indicator function associated with the sets~\eqref{eq:h sets} as:
\begin{equation}
\mathbf{1}_{\mathcal{S}_*}(\qe) \triangleq
\begin{cases}
1, & h_*(\qe) \leq 0 \\
0, & \text{otherwise}
\end{cases}
\label{eq:indicator-scalar}
\end{equation}
Hence, logical constraints~\eqref{eq:row-constr-conditional} can be reformulated as inequality constraints introducing \textit{big-M} constants \cite{bemporad_control_1999}:
\begin{equation}\label{eq:indicator-fun-constr}
\begin{cases}
     h_{\ell}^{\mathrm{row}}(\qD, \qDbar) \leq M_{\ell} \left(1-\mathbf{1}_{\mathcal{S}_s}\!(\qe) \mathbf{1}_{\mathcal{S}_{n^+}}\!(\bar{\qe})\right), \\
    h_{r}^{\mathrm{row}}(\qD, \qDbar) \leq M_{r} \left( 1- \mathbf{1}_{\mathcal{S}_s}\!(\qe) \mathbf{1}_{\mathcal{S}_{n^-}}\!(\bar{\qe}) \right),
\end{cases}
\end{equation}
where \( M_* \in \mathbb{R} : h_*^{\text{row}}(\qD, \qDbar) \leq M_* \ \forall\, \q,\bar{\q} \in \mathcal{Q},\ *\!\in\!\{\ell,r\} \).

We finally introduce the dynamics of \(\bar{\qe}\), which, together with the constraint functions in~\eqref{eq:indicator-fun-constr}, enable the formulation of a predictive optimal motion planning problem. The evolution of \(\bar{\qe}\) can be modeled as a \emph{Sample-and-Hold} system:
\begin{equation}
    \bar{\qe}_{k+1} = 
    \begin{cases}
        \bar{\qe}_k \quad \text{if } \qe \in \mathcal{S}_{s^+}, \\
        \qe_k \quad \text{if } \qe \notin \mathcal{S}_{s^+}.
    \end{cases}
\end{equation}
Using the indicator function:
\begin{equation}
    \bar{\qe}_{k+1} = \bar{\qe}_k \mathbf{1}_{S_{s^+}}\!(\qe_k)  + \qe_k \left(1-\mathbf{1}_{S_{s^+}}\!(\qe_k)\right).
    \label{eq:sample-and-hold-dyamics}
\end{equation}

It is worth noting that the right-of-way constraints~\eqref{eq:indicator-fun-constr} and the sample-and-hold dynamics~\eqref{eq:sample-and-hold-dyamics} are inherently discontinuous due to the presence of indicator functions, and therefore cannot be directly incorporated into optimization problems solved by traditional gradient-based methods. While smooth approximations of the indicator functions may be employed to recover differentiability, such approaches generally introduce approximation errors that may compromise the effectiveness of the proposed formulation.
To avoid this issue, the next section presents an exact encoding of such constraints by leveraging the Mixed Logical Dynamical modeling framework. 

\section{Regulation-Compliant Motion Planning} \label{sec:regulation-compliant-MPC}
Starting from the indicator function formulation, we leverage the MLD modeling framework to derive a formulation that integrates both continuous and integer variables---specifically binary variables---into constraints where such integer variables enter \emph{linearly}.
\subsection{MLD Right-of-Way Constraints}
For each function defined in~\eqref{eq:h-functions} we introduce a binary variable \(\delta_* \in \{0,1\}\) along with the corresponding logical implication:
\begin{equation}
     [\delta_{*} = 1] \iff [h_{*}(\qe) \leq 0 ] \quad \forall\, * \in \mathcal{I}_{sn}.
     \label{eq:logical-implications}
\end{equation}
This relation can be encoded via the following set of mixed-integer inequalities:
\begin{equation}
    \begin{cases}
        h_{*}(\qe) \leq M_{*}(1-\delta_{*}), \\
        h_{*}(\qe) \geq \epsilon + (m_{*}-\epsilon)\delta_{*},
    \end{cases}
    \quad \forall\, * \in \mathcal{I}_{sn}
    \label{eq:MI-logical-implications}
\end{equation}
where \(M_{*},m_{*} \in \mathbb{R}\) are chosen such that \( m_* \leq h_*(\qe) \leq M_* \ \forall\, \qe \in \mathcal{Q}^2 \), and \(\epsilon\) is a small tolerance introduced for numerical robustness~\cite{bemporad_control_1999}. Note that, with a slight abuse of notation, the functions \(h_*(\cdot)\) in~\eqref{eq:logical-implications} and~\eqref{eq:MI-logical-implications} are evaluated at \(\bar{\qe}\) for \(*\in\{n^+,n^-\}\), in accordance with Def.~\ref{def:row constr-conditional}.

Notably, the product of two indicator functions translates into a logical AND condition. We then introduce the binary variables \(\delta_s \triangleq \delta_{s^+ \wedge s^-} \), \(\delta_{\ell} \triangleq\delta_{s \wedge n^+} \), and \(\delta_r \triangleq \delta_{s \wedge n^-} \), where we adopt the notation \(\delta_{a \wedge b} \triangleq \delta_a \wedge \delta_b\). The logical AND condition is enforced through the following inequalities:
\begin{equation}
\begin{cases}
    \delta_{a \wedge b} \leq \delta_a, \\
    \delta_{a \wedge b} \leq \delta_b, \\
    \delta_a + \delta_b \leq \delta_{a \wedge b} + 1,
\end{cases}
\quad \forall\, (a,b) \in \mathcal{I}_{\wedge}
\label{eq:logical-AND}
\end{equation}
with \(\mathcal{I}_{\wedge} \triangleq \{(s^+,s^-),\ (s,n^+),\ (s,n^-)\}\).
The constraints in~\eqref{eq:indicator-fun-constr} can now be rewritten using the binary variables \(\delta_{\ell}\) and \(\delta_{r}\) as follows:
\begin{equation*}
\begin{cases}
    h_{\ell}^{\mathrm{row}}(\qD, \qDbar) \leq M_{\ell} \left(1-\delta_{\ell}\right), \\
    h_{r}^{\mathrm{row}}(\qD, \qDbar) \leq M_{r} \left( 1- \delta_{r} \right).
\end{cases}
\end{equation*}

Note that the \(\min(\cdot)\) operation in~\eqref{eq:effective-granted-space} can be explicitly modeled within the MLD framework.  
To this end, we introduce binary variables \(\delta_{g\ell}\) and \(\delta_{gr}\), as well as auxiliary continuous variables \(z_{g\ell}\) and \(z_{gr}\). The following mixed-integer constraints enforce \(z_{g\ell} = \Delta_{g\ell}(\qDbar)\) and \(z_{gr} = \Delta_{gr}(\qDbar)\):
\begin{equation}
    \begin{cases}
        \Delta_{g}^{\mathrm{row}} - M_{g*}(1-\delta_{g*}) \leq z_{g*} \leq \Delta_{g}^{\mathrm{row}},\\
        \Delta_{*}(\qDbar) - M_{g*}\delta_{g*} \leq z_{g*} \leq \Delta_{*}(\qDbar),
    \end{cases}
    \quad \forall\, * \in \{\ell, r\}
    \label{eq:min-logic-form}
\end{equation}
where \(M_{g*} \in \mathbb{R} : |\Delta_{g}^{\mathrm{row}} - \Delta_{*}(\q)| \leq M_{g*} \ \forall\, \q \in \mathcal{Q}\).
The MLD form of constraints~\eqref{eq:indicator-fun-constr} becomes:
\begin{equation}
    \begin{cases}
        -\Delta_{\ell}(\qD) + z_{g\ell} \leq M_{\ell} \left(1-\delta_{\ell}\right),\\
        -\Delta_{r}(\qD) + z_{gr} \leq M_{r} \left( 1- \delta_{r} \right). \\
    \end{cases}    
    \label{eq:MLD-row}
\end{equation}
Finally, we collect all constraints introduced in~\eqref{eq:MI-logical-implications}, \eqref{eq:logical-AND}, \eqref{eq:min-logic-form}, and~\eqref{eq:MLD-row}, and express them in compact form by moving all terms to the left-hand side:
\begin{equation}
    \boldsymbol{\gamma}^{\mathrm{row}}\left(\qe, \bar{\qe}, \boldsymbol{\delta}^{\mathrm{row}}, \z^{\mathrm{row}}\right) \leq \mathbf{0},
    \label{eq:gamma^row}
\end{equation}
where the stacked binary and continuous auxiliary variables are defined as:
\begin{align}
        \boldsymbol{\delta}^\mathrm{row} & \triangleq \begin{bmatrix} \{\delta_{*}\}_{*\in \mathcal{I}_{sn}} & \delta_s & \delta_{g\ell} & \delta_{gr} & \delta_{\ell} & \delta_{r} \end{bmatrix}^{\top}, \label{eq:delta-vector}\\
        \z^{\mathrm{row}} & \triangleq \begin{bmatrix} z_{g\ell} \quad z_{gr} \end{bmatrix}^{\top}.
\end{align}

\subsection{MLD Sample-and-Hold Dynamics}
The dynamics of \(\bar{\qe}\) in~\eqref{eq:sample-and-hold-dyamics} can be rewritten using the previously introduced binary variable \(\delta_{s^+}\):
\begin{equation}
    \begin{aligned}
    \bar{\qe}_{k+1} &= \bar{\qe}_{k}\delta_{s^+,k} + \qe_k(1-\delta_{s^+,k}) \\
    &= (\bar{\qe}_k - \qe_k)\delta_{s^+,k} + \qe_k. \\
    \end{aligned}
\end{equation}
Here, we replace the bilinear term involving the binary variable with a vector of auxiliary continuous variables \(\mathbf{z}_{\mathrm{sh}} \in \mathbb{R}^4\). The following mixed-integer constraints enforce \(\mathbf{z}_{\mathrm{sh}}=(\bar{\qe} - \qe)\, \delta_{s^+}\):
\begin{align}
\begin{cases}
     \mathbf{m}_{\mathrm{sh}}\delta_{s^+} \leq \mathbf{z}_{\mathrm{sh}}  \leq \mathbf{M}_{\mathrm{sh}}\delta_{s^+}, \\
    (\bar{\qe} - \qe) - \mathbf{M}_{\mathrm{sh}}(1-\delta_{s^+}) \leq \mathbf{z}_{\mathrm{sh}}, \\
    (\bar{\qe} - \qe) - \mathbf{m}_{\mathrm{sh}}(1-\delta_{s^+}) \geq \mathbf{z}_{\mathrm{sh}},
\end{cases}
\label{eq: sample & hold MLD constr}
\end{align}
where \(\mathbf{m}_{\mathrm{sh}},\, \mathbf{M}_{\mathrm{sh}} \in \mathbb{R}^4\) are constant vectors satisfying \(\mathbf{m}_{\mathrm{sh}} \leq (\bar{\qe} - \qe)\leq \mathbf{M}_{\mathrm{sh}} \ \forall\, \qe,\bar{\qe} \in \mathcal{Q}^2\).
Defining the function \(\mathbf{f}_{\mathrm{sh}}(\qe_k, \z_{\mathrm{sh},k}) \triangleq \qe_k + \z_{\mathrm{sh},k}\),
the MLD sample-and-hold system is described by:
\begin{equation}
\begin{aligned}
    &\bar{\qe}_{k+1} = \mathbf{f}_{\mathrm{sh}}(\qe_k, \z_{\mathrm{sh},k}),\\
    &\boldsymbol{\gamma}^{\mathrm{sh}}(\qe_k,\bar{\qe}_k,\delta_{s^+,k},\z_{\mathrm{sh},k}) \leq \mathbf{0},
\end{aligned}    
\end{equation}
where \(\boldsymbol{\gamma}^{\mathrm{sh}}(\cdot)\) collects constraints~\eqref{eq: sample & hold MLD constr} along with constraints~\eqref{eq:MI-logical-implications} for \(* = s^+\). 

\subsection{MLD Collision Avoidance Constraints}
Let \(\Delta_s^{\mathrm{CA}}\) and \(\Delta_n^{\mathrm{CA}}\) be longitudinal and lateral safety margins, accounting for vehicles dimensions, racing line curvature, and maximum orientation misalignment. Introduce the following scalar functions:
\begin{subequations}
        \begin{align}
            h_{s^{\pm}}^{\mathrm{CA}}(\qe) &\triangleq \pm s_{\mathrm{D}} \mp s_{\mathrm{A}} + \Delta_s^{\mathrm{CA}},\\
            h_{n^{\pm}}^{\mathrm{CA}}(\qe) &\triangleq \pm n_{\mathrm{D}} \mp n_{\mathrm{A}} + \Delta_n^{\mathrm{CA}}.
        \end{align}
\end{subequations}
Then collision-avoidance is ensured if:
\begin{equation}
    h_*^{\mathrm{CA}}(\qe) \leq 0 \quad \text{for at least one } * \in \mathcal{I}_{sn}.
\end{equation}
Introducing the vector of binary variables \(\boldsymbol{\delta}^\mathrm{CA}\triangleq\{\delta^\mathrm{CA}_*\}_{*\in\mathcal{I}_\mathrm{CA}}\):
\begin{equation}
    \begin{cases}
        h_*^\mathrm{CA}(\qe) \leq M_*^\mathrm{CA}(1-\delta_*^\mathrm{CA}),\\
        \sum_{*\in \mathcal{I}_{sn}} \delta_*^\mathrm{CA} \geq 1
    \end{cases} \quad \forall\,*\in\mathcal{I}_{sn},
    \label{eq:CA-constraints}
\end{equation}
where \(M_*^\mathrm{CA} \in \mathbb{R} : h_{*}^{\mathrm{CA}}(\qe) \leq M_*^\mathrm{CA} \ \forall\, \qe \in \mathcal{Q}^2 \quad \forall\, *\in\mathcal{I}_{sn}\).
Finally, we collect the collision avoidance constraints~\eqref{eq:CA-constraints} in:
\begin{equation}
    \gamma^\mathrm{CA}(\qe,\boldsymbol{\delta}^\mathrm{CA})\leq 0.
\end{equation}

\subsection{Regulation-Compliant Model Predictive Control}
For notational convenience, we define \( (i, -i) \in \{\mathrm{A}, \mathrm{D}\} \) to denote a generic player--opponent pair, where \( i \) refers to the player under consideration (either the attacker or the defender), and \( -i \) denotes the corresponding opponent. Also for clarity, we replace the joint notation \( \qe \) with explicit arguments \( (\x^i, \x^{-i}) \), representing the individual player states, given that the mapping from the state \(\x^i\) to the position \(\q^i\) is straightforward.

For each player \(i\), let \( \mathcal{X}^i \subseteq \mathbb{R}^{n^i} \) and \( \mathcal{U}^i \subseteq \mathbb{R}^{m^i} \) denote the admissible state and control input sets, respectively. 
We define \( \mathbf{r}_x^i : \mathcal{X}^i \rightarrow \mathcal{X}^i \) and \( \mathbf{r}_u^i : \mathcal{X}^i \rightarrow \mathcal{U}^i \) as the state dependent reference state and control input, respectively, and assume that \( \mathbf{r}_x^i,\; \mathbf{r}_u^i \in C^2 \), i.e., both functions are twice continuously differentiable.
We consider the stage and terminal cost functions:  
\begin{align}
    \ell^i(\x_k^i, \u_k^i) &\triangleq 
    \left\| \x_k^i - \mathbf{r}_x^i(\x_k^i) \right\|_{Q^i}^2 +
    \left\| \u_k^i - \mathbf{r}_u^i(\x_k^i) \right\|_{R^i}^2, \label{eq:stage-cost} \\
    \ell_f^i(\x_N^i) &\triangleq 
    \left\| \x_N^i - \mathbf{r}_x^i(\x_N^i) \right\|_{P^i}^2, \label{eq:terminal-cost}
\end{align}
where \( Q^i \succeq 0 \), \( R^i \succ 0 \), and \( P^i \succeq 0 \) are the weighting matrices for the state, input, and terminal cost, respectively.
The total cost over the prediction horizon \(N\) is then:
\begin{equation}
    J^i(X^i,U^i) \triangleq \sum_{k=0}^{N-1} \ell^i(\x_k^i, \u_k^i) + \ell_f^i(\x_N^i),
\end{equation}
where \(X^i \triangleq \{ \x_k^i \}_{k=0}^{N}\) and \(U^i \triangleq \{ \u_k^i \}_{k=0}^{N-1}\).
Let \( \xe \) denote the state of a vehicle motion model in the Frenet frame, with dynamics \( \xe_{k+1} = \mathbf{f}_{\chi}(\xe_k, \u_k) \).
We define the state, binary, and auxiliary variables for each agent as:
\begin{equation*}
\begin{aligned}
    \xA &\triangleq \xeA, 
    &\qquad
    \xD &\triangleq \begin{bmatrix}
        \xeD & \bar{\qe}
    \end{bmatrix}^{\top}, \\
    \dA &\triangleq \boldsymbol{\delta}^{\mathrm{CA}}, 
    &\qquad
    \dD &\triangleq \boldsymbol{\delta}^{\mathrm{row}}, \\
    \zA &\triangleq \varnothing, 
    &\qquad
    \zD &\triangleq \begin{bmatrix}
        \z^{\mathrm{row}} & \z_{\mathrm{sh}}
    \end{bmatrix}^{\top}.
\end{aligned}
\end{equation*}
The attacker and defender states evolve according to:
\begin{equation}
\begin{aligned}
    \xAk{k+1} &= \mathbf{f}^{\mathrm{A}}(\xA_k, \uA_{k}) \triangleq
    \mathbf{f}_{\chi}(\xA_k, \uA_k), \\
    \xD_{k+1} &= \mathbf{f}^{\mathrm{D}}(\xD_k, \uD_{k}, \zD_k, \xA_k) 
    \triangleq \begin{bmatrix} \mathbf{f}_{\chi}(\xD_k, \uD_k) \\
    \mathbf{f}_{\mathrm{sh}}(\xD_k, \zDk{k}, \xAk{k}) \end{bmatrix}.
\end{aligned}
\end{equation}
Each player is subject to stage-wise state-input constraints and a terminal state constraint:
\begin{equation}
    \begin{aligned}
    \mathbf{g}^{i}_{x,u}(\x^{i}_{k},\u^{i}_k) \leq \mathbf{0},\\
        \mathbf{g}^{i}_{N}(\x^{i}_{N}) \leq \mathbf{0}.
    \end{aligned}
\end{equation}
Finally, the overtaking regulations are encoded via coupling constraints:
\begin{equation}
\begin{aligned}
    \gammaA(\xA, \dA, \xD) &\triangleq \boldsymbol{\gamma}^{\mathrm{CA}}(\xA, \dA, \xD), \\
    \gammaD(\xD, \dD, \zD, \xA) &\triangleq 
    \begin{bmatrix}
        \boldsymbol{\gamma}^{\mathrm{sh}}(\xD,\dD,\zD,\xA) \\ \boldsymbol{\gamma}^{\mathrm{row}}(\xD, \dD, \zD)
    \end{bmatrix}. \label{eq:MI-ineq-gamma}
\end{aligned}
\end{equation}
We now formally define the player-specific MPC problem that incorporates regulation constraints.
\begin{definition}[Regulation-Compliant MPC (RC-MPC)] \label{def:regulation-compliant-MPC}
The \emph{Regulation-Compliant MPC (RC-MPC)} formulation that enforces overtaking regulation compliance for the player \(i\) is defined as\footnote{The time instant \( t \) is omitted from prediction variables for simplicity, assuming all quantities are defined relative to the current time.}:
    \begin{subequations}
    \begin{align}
        \min_{X^i,U^i,\Delta^i,Z^i} &J^i(X^i,U^i) \\
        \text{s.t. } &\x^{i}_{k+1} = \mathbf{f}^i(\x^{i}_k, \u^{i}_k, \z^{i}_k, \x^{-i}_k), &&k\in \mathbb{I}_0^{N-1} \label{eq:RC-MPC-dynamics}\\
        &\x^{i}_{0} = \x^{i}(t), \label{eq:RC-MPC-initial-cond}\\
        &\mathbf{g}^{i}_{N}(\x^{i}_{N}) \leq \mathbf{0}, \label{eq:RC-MPC-terminal}\\
        &\mathbf{g}^{i}_{x,u}(\x^{i}_{k},\u^{i}_k) \leq \mathbf{0}, &&k\in \mathbb{I}_0^{N-1} \label{eq:RC-MPC-state-input-constr}\\
        &\boldsymbol{\gamma}^{i}(\x^{i}_k, \boldsymbol{\delta}^{i}_k, \z^{i}_{k}, \x^{-i}_k) \leq \mathbf{0}, &&k\in \mathbb{I}_0^{N} \label{eq:RC-MPC-MLD-constr}
    \end{align}
    \label{prob:RC-MPC}
    \end{subequations}
    where the optimization variables are \( X^{i} \triangleq \{ \x^{i}_{k} \}_{k=0}^{N}\), \(U^{i} \triangleq \{ \u^{i}_k \}_{k=0}^{N-1}\), \( \Delta^{i} \triangleq \{ \boldsymbol{\delta}^{i}_k \}_{k=0}^{N}\), \( Z^{i} \triangleq \{\z^{i}_k\}_{k=0}^{N}\).
\end{definition}

\begin{remark}
    Problem~\eqref{prob:RC-MPC} is a Mixed-Integer Nonlinear Program (MINLP), where (i) the binary variables appear only in the inequality constraints and enter linearly, and (ii) the cost function is assumed to be twice continuously differentiable.  
    In general, solving MINLPs is computationally challenging. To address this, we adopt a Mixed-Integer Sequential Quadratic Programming (MI-SQP) approach~\cite{quirynen_sequential_2021}.  
    Analogous to classical SQP, the MI-SQP method sequentially computes and solves Mixed-Integer Quadratic Program (MIQP) approximations of the original problem until convergence to a (local) minimizer.
\end{remark}

\section{Regulation-Aware Game-Theoretic Motion Planning} \label{sec:regulation-aware-GTMP}
In this section, we leverage the RC-MPC framework to model the interactions of the two vehicles during an overtaking maneuver.  
Specifically, we focus on the design of a game-theoretic motion planner that brings predictive reasoning capabilities over the regulation constraining the defender to enhance the effectiveness of the attacker's overtaking strategies.

The proposed interaction is modeled as a non-cooperative, two-player, finite-horizon, differential game with player-specific coupling constraints. 
\begin{definition}[Regulation-Constrained Racing Game]
We define the \emph{Regulation-Constrained Racing Game} at time \( t \) as:
\begin{equation}
\mathcal{G}(t) \triangleq 
\left\langle 
\mathcal{I},\,
\mathcal{P}^{i}(\x^{i}(t), X^{-i}(t))
\right\rangle_{i \in \mathcal{I}},
\label{eq:game-tuple}
\end{equation}
where:
\begin{itemize}
    \item \( \mathcal{I} \triangleq \{\mathrm{A}, \mathrm{D}\} \) is the set of players (attacker and defender);
    \item \( \x^{i}(t) \) is the state of agent \( i \) at current time  \(t\);
    \item \( X^{-i}(t)\) is the predicted state trajectory of the opponent over the horizon \( N \);
    \item \( \mathcal{P}^{i}(\x^{i}(t), X^{-i}(t)) \) is the RC-MPC problem~\eqref{prob:RC-MPC} solved by agent \( i \), initialized at \( \x^{i}(t) \), and conditioned on \( X^{-i}(t) \).
\end{itemize}
\end{definition}

To characterize the outcome of the game \( \mathcal{G} \), we adopt the solution concept of a Generalized Nash Equilibrium, where each player optimizes its own trajectory over a finite horizon, subject to \emph{player-specific} constraints that may depend on the opponent's strategy.
\begin{definition}[Generalized Nash Equilibrium (GNE)]
Let the strategy of player \( i \in \mathcal{I} \) be defined as \( w^i \triangleq (X^i, U^i, Z^i, \Delta^i) \), and let the feasible set be given by
\begin{equation}
    W^i(w^{-i}) \triangleq \left\{ w^i \mid w^i \text{ satisfies } \eqref{eq:RC-MPC-dynamics} \text{--} \eqref{eq:RC-MPC-MLD-constr} \right\}.
\end{equation}
A strategy pair \( \{w^{i\star}\}_{i \in \mathcal{I}} \) is a \emph{Generalized Nash Equilibrium} of the game \( \mathcal{G} \) in~\eqref{eq:game-tuple} if, for each player \( i \in \mathcal{I} \),
\begin{equation}
    w^{i\star} \in \mathrm{BR}^i(w^{-i\star}),
\end{equation}
where the \emph{best-response mapping} is defined as
\begin{equation}
    \mathrm{BR}^i(w^{-i}) \triangleq \arg\min_{w^i} J^i(w^i) \quad \text{s.t.} \quad w^i \in W^i(w^{-i}).
\end{equation}
Equivalently, a GNE is a fixed point of the collective best-response mapping.
\end{definition}

\begin{remark}
The regulation-constrained racing game is inherently asymmetric due to the player-specific regulations, which assign different coupling constraints to each player. 
Although both agents in reality are interested in avoiding collisions, this asymmetry is intended to model a reasoning framework where the regulation assigns greater responsibility for safety to the attacker. 
From a safety perspective, including collision avoidance constraints in the defender’s problem would be redundant: any feasible strategy pair inherently prevents collisions for \emph{both} agents, provided such a strategy pair exists.
On the other hand, enforcing collision avoidance also on the defender could admit GNEs in which the attacker exploits the defender's reactive behavior, for example by forcing it off the racing line. This may promote overly aggressive or unfair behavior, which motivates our choice to assign collision avoidance constraints only to the attacker.
Finally, we highlight that ensuring the existence of a GNE to this asymmetric and non-convex game is nontrivial and depends critically on the proper design of the regulatory framework.
\end{remark}

To compute an approximate solution to the GNE, we adopt an Iterative Best Response (IBR) scheme~\cite{facchinei_generalized_2010}, wherein each agent \( i \in \mathcal{I} \) sequentially updates its strategy as \( w^{i, (\ell+1)} \in \mathrm{BR}^i(w^{-i, (\ell)})\), assuming the opponent’s strategy \( w^{-i, (\ell)} \) is fixed.
The process is repeated until convergence to a fixed point of the best-response mapping (i.e. a GNE) or until a maximum number of iterations is reached.
Despite the difficulty of analyzing existence and convergence properties~\cite{facchinei_decomposition_2011}, these methods remain widely used in practice to approximate GNEs~\cite{wang_game-theoretic_2021,williams_best_2018,spica_real-time_2020}. 

We now formalize the strategy computed by the attacker.
\begin{definition}[Regulation-Aware Game-Theoretic Planner (RA-GTP)]
We define the \emph{Regulation-Aware Game-Theoretic Planner (RA-GTP)} for the attacker as the strategy \( w^{\mathrm{A}\star}(t)\) obtained from the IBR approximation of the GNE of the game \( \mathcal{G}(t) \), applied in a receding horizon fashion.
\end{definition}
Although the RA-GTP is defined for the attacker in this work, the formulation extends naturally to any player.

\section{Simulation Studies} \label{sec:simulation}
The simulation studies are conducted on the Autodromo di Modena racetrack layout. Both the attacker and the defender are assumed to follow the same racing line.  
Each vehicle follows a time-optimal speed profile subject to bounds on longitudinal and lateral accelerations. To encourage overtaking maneuvers, the attacker is assigned a reference speed profile with higher longitudinal and lateral accelerations compared to the defender.

The motion planning problems for both the attacker and the defender are formulated over a prediction horizon of \( N = 20 \) steps, with a sampling time of \( T_s = \SI{0.05}{s} \).
Both planning schemes rely on the discretized kinematic bicycle model in Appendix~\ref{sec:appendix}.
The right-of-way and collision avoidance thresholds are summarized in Table~\ref{tab:parameters}.
\begin{table}[t]
\centering
\caption{Right-of-Way and Collision Avoidance Thresholds}
\label{tab:parameters}
\begin{tabular}{@{}ll|ll@{}}
\toprule
\textbf{Parameter} & \textbf{Value} & \textbf{Parameter} & \textbf{Value} \\
\midrule
\(\Delta_s^{\mathrm{row}}\) & \(2.0 \times\) car length     & \(\Delta_s^{\mathrm{CA}}\) & \(1.5 \times\) car length \\
\(\Delta_n^{\mathrm{row}}\) & \(0.5 \times\) car width    & \(\Delta_n^{\mathrm{CA}}\) & \(1.5 \times\) car width \\
\(\Delta_g^{\mathrm{row}}\) & \(1.5 \times\) car width    &                             &                            \\
\bottomrule
\end{tabular}
\end{table}

The MI-SQP algorithm is implemented in Python, using CasADi~\cite{andersson_casadi_2019} for the computation of the quadratic approximations of the original MINLP, and GUROBI~\cite{gurobi_optimization_llc_gurobi_2024} as the MIQP solver. The IBR iterations are initialized using the solution from the previous planning step.

We conduct our simulation studies by comparing two planning schemes for the attacker: a baseline method and the proposed RA-GTP.
In both cases, the defender executes a RC-MPC strategy, which means it adheres to the rules. To this end, the defender is provided with the attacker's trajectory over the entire prediction horizon, while the attacker observes only the defender’s current state.
The baseline method solves the IBR using a defender model that does not account for regulation compliance. Equivalently, the baseline assumes a fixed prediction of the opponent's trajectory.

\begin{figure}[t]
    \centering
    \includegraphics[width=\columnwidth, clip=true, trim=0 0 0 0]{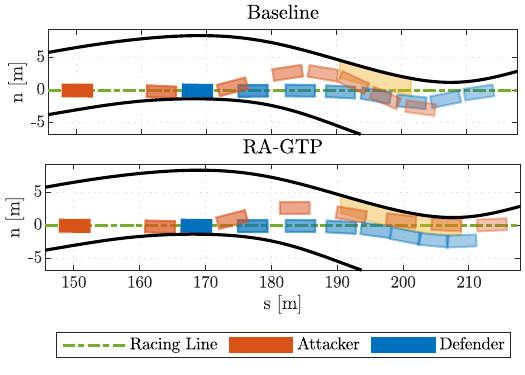}
    \caption{Overlay of attacker and defender positions during an overtaking attempt in the Frenet frame. The yellow area indicates the region prohibited to the defender due to right-of-way constraints. }
    \label{fig:overtake-sequence-frenet}
\end{figure}

The comparison highlights that RA-GTP achieves a higher success rate and more consistent overtaking maneuvers compared to the baseline approach. Specifically, the baseline is prone to abort the overtaking attempts in those regions of the track where the racing line move laterally across the track width, thus (erroneously) predicting the defender to shift sides and preventing the attacker from executing a safe pass.
In contrast, RA-GTP reasons over the defender’s right-of-way constraints and correctly predicts when lateral space will be granted (see Fig.~\ref{fig:overtake-sequence-frenet}).  
This enables the attacker to complete overtaking maneuvers in regions where the baseline would abort.

\begin{figure}[t]
    \centering
    \includegraphics[width=\columnwidth, clip=true, trim=0 0 0 0]{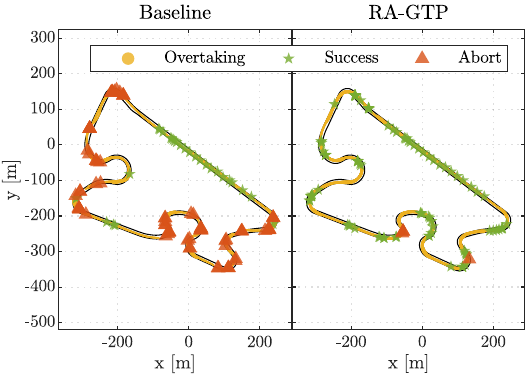}
    \caption{Overtaking outcomes across 100 simulations---ongoing, aborted, and completed maneuvers---show a 29~\% success rate for the baseline and 96~\% for RA-GTP (see the accompanying video\protect\footnotemark).}
    \label{fig:overtake-comparison}
\end{figure}

Figure~\ref{fig:overtake-comparison} shows the locations on the track where the attacker either successfully completed an overtaking maneuver (success) or failed the attempt (abort), under the two planning schemes.
The results are based on 100 test cases, where the attacker is initialized at a uniformly sampled position along the racing line, and the defender is placed a fixed distance ahead. Both vehicles start at their respective reference speed profiles.
Overtaking points are marked when the longitudinal gap between the vehicles drops below the right-of-way threshold \( \Delta_s^{\mathrm{row}} \).  
A point is labeled as a \emph{success} if the attacker subsequently moves ahead by at least \( \Delta_s^{\mathrm{row}} \), or an \emph{abort} if the attacker fails to complete the pass and the gap increases again beyond \( \Delta_s^{\mathrm{row}} \).
\footnotetext{\url{https://youtu.be/7Z6xG69eFhI}}

\section{Conclusion}
This work addressed the problems of regulation compliance and strategic overtaking in motion planning for autonomous racing.
We proposed a Regulation-Compliant Model Predictive Control (RC-MPC) formulation that enforces right-of-way rules and collision avoidance using Mixed Logical Dynamical constraints.  
Building on this, we introduced the Regulation-Aware Game-Theoretic Planner (RA-GTP), which enables the attacker to reason over the defender’s regulation-constrained behavior via an Iterative Best Response approximation of a Generalized Nash Equilibrium.

Simulation results demonstrate that RA-GTP achieves a higher overtaking success rate compared to baseline methods based on non-interactive opponent models.  
In particular, the proposed planner successfully exploits rule-compliant behavior to execute overtakes in scenarios where fixed-prediction approaches are prone to unnecessary aborts.  
These results highlight the importance of integrating regulatory awareness into motion planning to reduce conservatism and enable strategic reasoning.

Future work will focus on a deeper analysis of the asymmetric game setting, including the impact of player-specific coupling constraints on the existence of solutions to the GNEP.
We also plan to extend the framework to scenarios in which both agents leverage game-theoretic reasoning, and to assess the effectiveness and computational complexity of the proposed approach through real-world experiments.


\section*{APPENDIX}\label{sec:appendix}
Kinematic‑bicycle model (CoG reference) in Frenet coordinates:
\begin{equation}
\begin{aligned}
\dot{s}      &= \frac{v\cos\bigl(e_{\psi}+\beta\bigr)}{1-\kappa(s)\,n}, \\
\dot{n}      &= v\sin\bigl(e_{\psi}+\beta\bigr), \\
\dot{e}_{\psi} &= \frac{v}{L}\cos\beta\,\tan\delta - \kappa(s) \dot{s}, \\
\dot{v}      &= a, \\
\dot{\delta} &= \omega.
\end{aligned}
\label{eq:kinematic-frenet}
\end{equation}
Side‑slip (kinematic) at the CoG:
\(\beta = \arctan \Bigl(\tfrac{l_r}{L}\tan\delta\Bigr).\)



\bibliographystyle{IEEEtran_DOI}
\bibliography{IEEEabrv,references}


\end{document}